\documentclass[twocolumn,showpacs,showkeys,preprintnumbers,amsmath,amssymb]{revtex4}

\usepackage{graphicx}
\usepackage{dcolumn}
\usepackage{bm}

\begin{document}


\title{Statistics of Extreme Values in Time Series with  
Intermediate-Term Correlations}   

\author{Cecilia Pennetta}
\affiliation{   
Dipartimento di Ingegneria dell'Innovazione, Universit\`a del Salento
and CNISM, \\Via Arnesano, Edificio "Stecca", I-73100 Lecce, Italy}  

\begin{abstract}  
It will be discussed the statistics of the extreme values in time series 
characterized by finite-term correlations with non-exponential decay.  
Precisely, it will be considered the results of numerical analyses concerning 
the return intervals of extreme values of the fluctuations of resistance 
and defect-fraction displayed by a resistor with granular structure in a 
nonequilibrium stationary state. The resistance and defect-fraction are  
calculated as a function of time by Monte Carlo simulations using a  
resistor network approach. It will be shown that when the auto-correlation 
function of the fluctuations displays a non-exponential and non-power-law 
decay, the distribution of the return intervals of extreme values is a 
stretched exponential, with exponent largely independent of the threshold. 
Recently, a stretched exponential distribution of the return intervals of 
extreme values has been identified in long-term correlated time series by 
Bunde et al. (2003) and Altmann and Kantz (2005). Thus, the present results  
show that the stretched exponential distribution of the return  
intervals is not an exclusive feature of long-term correlated time series. 
\end{abstract}  


\maketitle

 
\section{INTRODUCTION}  
\label{sect:intro}    
The puzzling questions posed by the observations of climate changes  
\cite{christensen,mudelsee,tebaldi,perry,goswami,ice_1,ice_3} have 
recently provided further motivations to the investigation of extreme events 
and dynamics \cite{gumbel,storch,kotz,sornette,racz,salvadori}. In particular,
the statistics of the return times (RTs) of extreme values is one of the 
subjects that have recently attracted the attentions of many authors 
\cite{schmitt,bunde_physa2003,bunde_prl2005,kantz_2005,upon05,pen_epj}. 
Return times (or intervals) $r_q$, of the extreme values of a fluctuating 
quantity $x(t)$, associated with the overcoming of a given threshold $q$, 
can be defined \cite{storch,bunde_physa2003} as the time intervals between 
two consecutive occurrences of the condition $x(t)>q$ . In the case of  
uncorrelated events it is easy to show that the RTs are exponentially 
distributed \cite{storch,kotz,bunde_physa2003}. Exponential distributions 
describe the statistics of return intervals of prices in financial markets, 
wind speed data or daily precipitations in a given place for the same time 
windows \cite{storch,bunde_physa2003,kantz_2005}. 
 
On the other hand, in recent years it has been realized that several  
other important examples of time series display long-term correlations  
\cite{bunde_physa2003,bunde_prl2005,kantz_2005}. This is the case  
of physiological data (heartbeats \cite{bunde_prl2000,ashkenazy} and  
neuron spikes \cite{davidsen}), hydro-meteorological records (daily  
temperatures \cite{bunde_physa2003,kantz_2005,koscielny_prl98,salvadori}), 
geophysical or astrophysical data (occurrence of earthquakes  
\cite{bak,corral} or solar flares \cite{boffetta}), internet traffic  
\cite{bunde_physa2003} and stock market volatility \cite{kantz_2005,liu} 
records. Long-term correlated time series are characterized by an  
auto-correlation function, $C_x(s)$, decaying as a power-law:  
\begin{equation} 
C_x(s) = <x_ix_{i+s}> = { 1 \over N-s} \sum_{i=1}^{N-s} x_ix_{i+s} 
\sim s^{-\gamma_x}  \label{long_corr}   
\end{equation}   
where the correlation exponent $\gamma_x$ ranges between 0 and 1  
(normalized series $\{x_i\}_{i=1,...N}$ with zero average and unit variance 
are considered). Therefore, in these conditions, the mean correlation time 
$\tau$ diverges ($\tau$ is defined as the integral over $s$ of $C_x$). The 
effect of long-terms correlations on the RT statistics has been recently 
studied by Bunde et al. \cite{bunde_physa2003,bunde_prl2005} and by Altmann 
and Kantz \cite{kantz_2005}. Both these studies were performed on the  
ground of numerical calculations on Gaussian and long-term correlated time 
series generated by the Fourier transform technique and by imposing a  
power-law decay of the power spectrum \cite{makse}. The main conclusions 
of Bunde at al. were the following. i) The mean return interval $R_q$ is 
unchanged by the presence of long-terms correlations. ii) The distribution 
of the RTs becomes stretched exponential: 
%
\begin{equation} 
P_q(r) = a/R_q \ \exp\bigl[ -\bigl(b \ \ r / R_q \bigr)^{\gamma} \bigr]  \label{stretch} 
\end{equation}    
%
where the exponents $\gamma$ and $\gamma_x$ were found to be equal. 
iii) The RTs themselves are long-term correlated with a correlation  
exponent $\gamma' \approx \gamma_x$. Actually, as noted by Altmann and Kantz 
\cite{kantz_2005}, the statement i) is true in general and can be identified 
with Kac's lemma \cite{kac} introduced in the context of dynamical  
systems while the statement iii) is true only in very special conditions 
\cite{kantz_2005}. Moreover, it must be noted that the results ii) and iii) 
only apply to linear time series (i.e. to series whose properties are  
completely defined by the power spectrum and by the probability density) 
\cite{kantz_2005,bunde_prl2005}. However, apart from this restriction,  
the stretched exponential distribution of the RTs can be considered a  
general feature in presence of long-term correlations in a time series  
\cite{kantz_2005,bunde_prl2005}. This feature has important consequences 
on the observation of extreme events: indeed it implies a strong enhancement 
of the probability of having return intervals well below and well above  
$R_q$, in comparison with the occurrence of extreme events in an uncorrelated 
time series. Furthermore, it must be noted \cite{kantz_2005} that the  
distribution in Eq.~(\ref{stretch}) only depends on the parameter $\gamma$, 
being $a$ and $b$ function of $\gamma$.  
 
Here it will be considered the effect on the distribution of the RTs of  
the presence of finite-term correlations with non-exponential decay, a  
situation which can occur in systems which are approaching criticality,  
where intermediate behaviors with non-exponential and non-power-law decay 
of correlations can emerge \cite{sornette}. Thus, in the next section  
it will be analyzed the RT statistics of extreme values of the resistance 
ad defect-fraction fluctuations displayed by a resistor with granular  
structure in a nonequilibrium stationary state \cite{pen_msn}. 
 
\section{METHOD AND RESULTS }   
The time series analyzed consist in the fluctuations of resistance and 
defect-fraction of a thin resistor with granular structure, biased by an 
external current $I$ and in contact with a thermal bath at temperature $T$. 
The resistance values are calculated by using the multi-species resistor 
network (MSN) model \cite{pen_msn}. The MSN model describes a conducting 
film with granular structure as a resistor network made by different species 
of elementary resistors. Precisely, the resistor species are distinguished 
by their resistances and by their energies associated with thermally  
activated stochastic processes of breaking and recovery of the elementary 
resistors. The broken elementary resistors, named defects, correspond 
to resistors of resistance about $10^9$ higher than that of normal resistors. 
As a result of the competition between pairs of breaking and recovery 
processes for each species, the network can reach a stationary state whose 
properties will be dependent on the external conditions. The resistance and 
the defect fraction of the network are calculated by Monte Carlo simulations 
as a function of the time for different temperatures \cite{pen_msn}.  
All the details about the MSN model and its results can be found  
in Ref. \onlinecite{pen_msn}. It should be noted, however, that this model  
provides a correlation time of the resistance fluctuations dependent  
on the temperature: $\tau \sim (T-T^*)^{-\theta}$ with ${\theta}=2.7$,  
thus diverging for $T \rightarrow T^*$. As a result, the power  
spectral density of the resistance fluctuations shows a $1/f^{\alpha}$  
dependence of frequency, with ${\alpha} \approx 1$ for $T\leq T^*$.  
It should be also noted that the MSN model represents an extension  
to systems characterized by $1/f$ noise of a previous model existing 
in the literature, the stationary and biased resistor network (SBRN) model 
\cite{pen_pre,pen_physa,pen_prb}. In any case, it must be underlined that, 
apart from the specific system described by the MSN model, the method used 
here for generating the time series can be also viewed as a pure numerical 
algorithm for generating numerical series with different and tunable 
correlation properties. 
 
\begin{figure}  
   \begin{center}  
   \begin{tabular}{c}  
   \includegraphics[height=6cm]{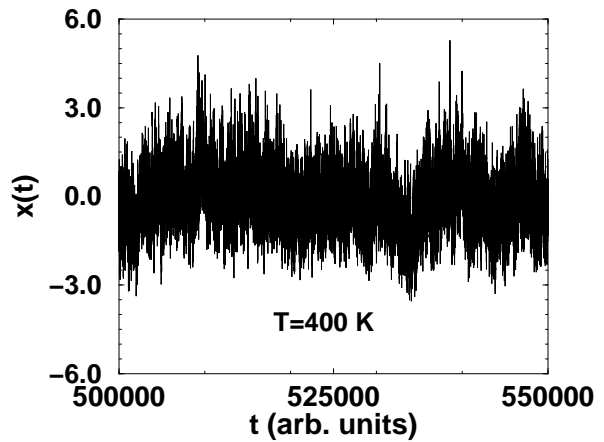}  
   \end{tabular}  
   \end{center}  
   \caption[signal]   
   {\label{fig:signal}  
   Normalized $x(t)$ signal (zero average and unit variance). Here the  
   original records consist of resistance fluctuations calculated at $400$ 
   K. The time is expressed in simulation steps.}  
\end{figure}   
 
\begin{figure}  
   \begin{center}  
   \begin{tabular}{c}  
   \includegraphics[height=6cm]{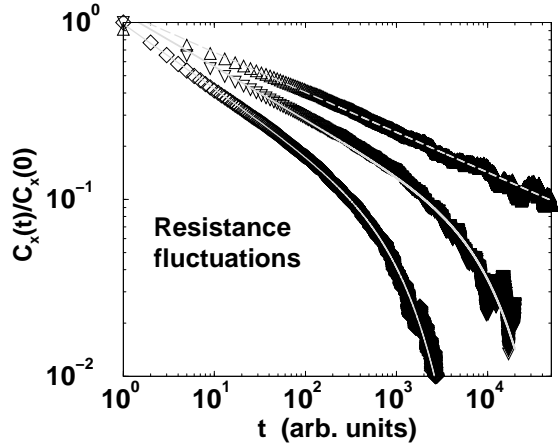}  
   \end{tabular}  
   \end{center}  
   \caption[correl_res]   
   {\label{fig:correl_res}  
   Auto-correlation functions of the resistance fluctuations calculated at 
   $300$ K (up triangles), $400$ K (down triangles) and $500$ K (diamonds). 
   The grey solid curves represent the best-fit with the function given in 
   Eq.~(\ref{mixed}), while the grey dashed curve shows the best-fit with 
   a power-law.} 
\end{figure}   
 
\begin{figure}  
   \begin{center}  
   \begin{tabular}{c}  
   \includegraphics[height=6cm]{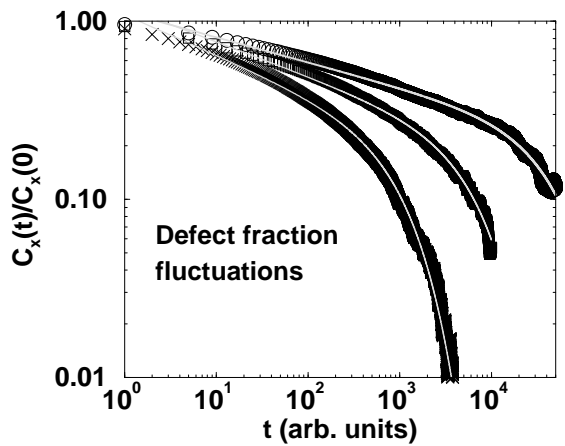}  
   \end{tabular}  
   \end{center}  
    \caption[correl_frac]   
   {\label{fig:correl_frac}  
   Auto-correlation functions of the defect-fraction fluctuations calculated 
   at $300$ K (up circles), $400$ K (squares) and $500$ K (crosses). 
   The grey solid curves represent the best-fit with the function given in 
   Eq.~(\ref{mixed}).} 
\end{figure}   
 
Long $R(t)$ and $p(t)$ time series (typically made of $1\div 2 \times 10^6$ 
records) have been generated and analyzed for different temperatures. Here 
$R$ indicates the resistance of the network expressed in Ohm and $p$ the 
defect-fraction, while the time $t$ is expressed in iteration steps. The 
analysis has been performed by considering normalized series:  
$x(t) \equiv (R(t)-<R>)/\sigma$, where $<R>$ is the average value of the network resistance and $\sigma$ the root-mean-square deviation from the  
average (a similar procedure has been adopted for the $p(t)$ series).  
The following quantities have been considered: the auto-correlation function, 
the PDF of the $x$ records, the return intervals $r_q$ of the extreme values 
for different threshold $q$ and their distribution $P_q(r)$ ($q$ is  
expressed in units of $\sigma$). All the results shown here are obtained 
by considering a network of size $75 \times 75$, biased by a current $I=1.0$A 
and made by 15 species of resistors. The details concerning the network can be found in Ref. \onlinecite{pen_msn}. 
 
Figure 1 displays a typical trend of $x(t)$. Only a small portion of the 
total number of records, $N=2 \times 10^6$, is shown. In this case $x(t)$ 
represents the normalized resistance fluctuations calculated at a  
temperature of $400$ K. The probability density of several $x(t)$ series 
has been also analyzed. It has been found that the PDF of the resistance fluctuations presents a significant non-Gaussianity that progressively  
increases at higher temperatures (going from $300$ to $500$ K)  
\cite{pen_msn,pen_physa,bramwell_nat}, while the PDF of the  
defect-fraction fluctuations is Gaussian at all the temperatures. 
  
Figure 2 reports the auto-correlation functions, $C_x(t)$,  
of three resistance fluctuation time series obtained for three  
different temperatures. Precisely, the three black curves  
correspond to a temperature of $300$, $400$ and $500$ K respectively going 
from the upper to the lower one. The grey dashed curve shows the best-fit 
to the $C_x$ data at $300$ K with a power-law of exponent $\gamma=0.22$, 
while the grey solid curves represent the best-fit to the data at $400$ and 
$500$ K with the function: 
\begin{equation} 
C_x(s)=C_0 t^{-h} \exp (-t/u)  \label{mixed}   
\end{equation}    
The fitting parameters are $C_0=1.23$, $h=0.30$ and $u=1.42 \times 10^4$ 
for the data at $400$ K and $C_0'=0.98$, $h'=0.36$ and $u'=1.51 \times 10^3$ 
for the data at $500$ K. Many other functions have been also considered  
for the best-fit of the $C_x$ data. However, it has been found that the 
function in Eq.~(\ref{mixed}) optimizes the best-fit procedure with the 
minimum numbers of fitting parameters. By using Eq.~(\ref{mixed}), it 
is possible to calculate the correlation times characterizing the resistance 
fluctuations: $\tau=1051$ at $400$ K and $\tau=148$ at $500$ K (see Ref. 
\onlinecite{pen_msn} for further details).  
Figure 3 shows the auto-correlation functions of the defect-fraction  
times series obtained at $300$, $400$ and $500$ K (black curves, again 
going from the upper to the lower one) and the best-fit to the $C_x$ data 
with Eq.~(\ref{mixed}) (grey solide curves).  
The fitting parameters are $C_0=1.12$, $h=0.14$ and $u=5.96 \times 10^4$ 
for the data at $300$ K, $C_0'=1.19$, $h'=0.19$ and $u'=8.17 \times 10^3$ 
for the data at $400$ K and $C_0''=1.08$, $h''=0.23$ and  
$u''=1.29 \times 10^3$ for the data at $500$ K. The correlation times are: 
 $\tau=1.40 \times 10^4$ at $300$ K, $\tau=1653$ at $400$ K and 
$\tau=306$ at $500$ K.  
 
\begin{figure}  
   \begin{center}  
   \begin{tabular}{c}  
   \includegraphics[height=6cm]{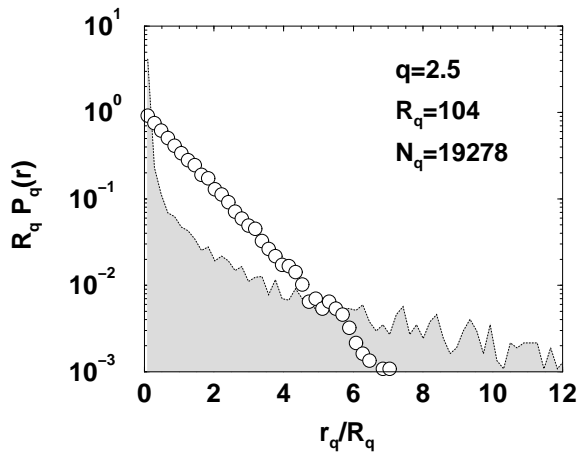}  
   \end{tabular}  
   \end{center}  
   \caption[isto]   
   {\label{fig:isto}  
   Distribution of the return intervals for values above the threshold  
   $q=2.5$ (dashed black curve at the border of the grey area) obtained  
   from the resistance fluctuations calculated at $300$ K (long-term  
   correlated records). For comparison, the distribution of the return  
   intervals for the same threshold value and obtained after  
   shuffling the data is shown by circles.} 
\end{figure}  
 
\begin{figure}  
   \begin{center}  
   \begin{tabular}{c}  
   \includegraphics[height=6cm]{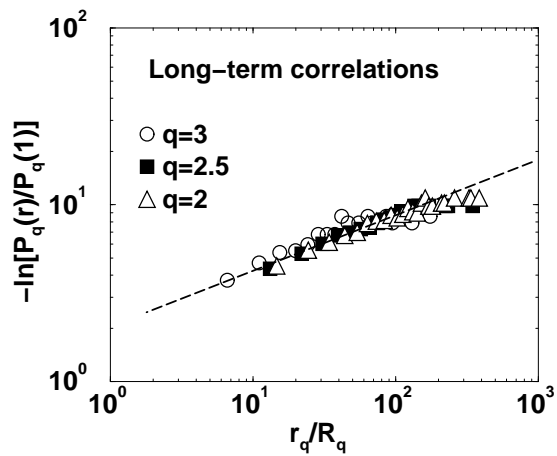}  
   \end{tabular}  
   \end{center}  
   \caption[dis_longterm]   
   {\label{fig:dis_longterm}  
   Double-logarithmic plot of the normalized probability density  
   of the return intervals for different threshold values $q$. The  
   $x(t)$ records are the same of Fig. 4, i.e. they present  
   long-term correlations. The slope of the dashed line is 0.3.} 
\end{figure}  
 
Let us start the analysis of the return intervals by considering first the 
RT distribution for extreme values of the resistance fluctuations at 
$300$ K. As pointed out by Fig. 2, in this case $C_x$ scales as a power-law 
over more than four orders of magnitude, indicating the existence of long-term 
correlations. Figure 4 reports the distribution of $r_q$ for $q=2.5$ 
in semi-logarithmic plot. Precisely, a normalized representation has been 
adopted for convenience, by reporting the product of mean return interval 
$R_q$ for the probability density $P_q(r)$ as a function of the ratio  
$r_q/R_q$. For comparison, the distribution of $r_q$ for the same threshold, 
obtained after shuffling the $x(t)$ records is also shown. 
Thus Fig. 4 highlights the typical effect of long-term correlations: 
strong enhancement of the probability of having return intervals well  
below and well above $R_q$, in comparison with the occurrence of extreme 
values in uncorrelated time series, as described in Refs.  
\onlinecite{bunde_physa2003,bunde_prl2005,kantz_2005}. Actually, the  
distribution of $r_q$ is found to be a stretched exponential, in  
agreements with the results of Refs. \onlinecite{bunde_physa2003,bunde_prl2005,kantz_2005}. 
This is shown in Fig. 5, which reports a double-logarithmic plot of the  
normalized probability density of the $r_q$ as a function of the ratio  
$r_q/R_q$ and for different thresholds. In this representation, the slope 
of the dashed line directly provides the exponent $\gamma$ in   
Eq.~(\ref{stretch}). As expected 
\cite{bunde_physa2003,bunde_prl2005,kantz_2005}, 
$\gamma =0.3 \approx \gamma_x$. 
 
\begin{figure}  
   \begin{center}  
   \begin{tabular}{c}  
   \includegraphics[height=6cm]{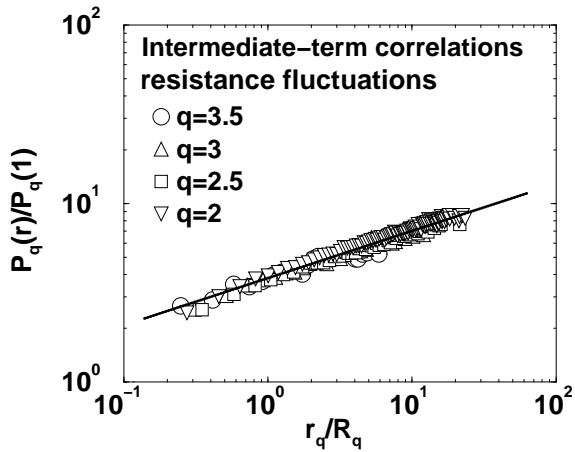}  
   \end{tabular}  
   \end{center}  
   \caption[dis_inter_res]   
   {\label{fig:dis_inter_res}  
   Double-logarithmic plot of the normalized probability density  
   of the return intervals for different threshold values $q$. In this  
   case the data (resistance fluctuations at $400$ K) are no more  
   long-term correlated but exhibit intermediate-term correlations. 
   The slope of the dashed line is 0.26.} 
\end{figure}  
 
\begin{figure}  
   \begin{center}  
   \begin{tabular}{c}  
   \includegraphics[height=6cm]{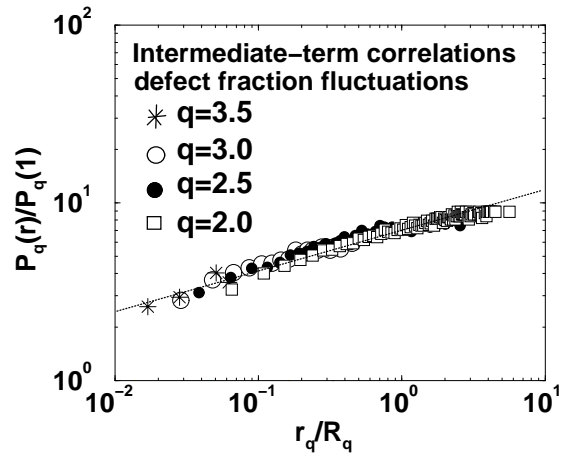}  
   \end{tabular}  
   \end{center}  
   \caption[dis_inter_frac]   
   {\label{fig:dis_inter_frac}  
   Double-logarithmic plot of the normalized probability density  
   of the return intervals for different threshold values $q$. Also in  
   this case the data (defect-fraction fluctuations at $400$ K) are  
   characterized by intermediate-term correlations. The slope of the  
   dashed line is 0.23.} 
\end{figure}  
 
Now, let us consider the RT distribution of extreme values for time series 
with finite correlation time and slow, non-exponential decay of correlations: 
a situation that can be called of "intermediate-term correlations". As shown 
by Figs. 2 and 3, the auto-correlation functions of the resistance  
fluctuations at $400$ and $500$ K (i.e. at $T>T^*$) and of the  
defect-fraction fluctuations at $300\div500$ K are well described by 
Eq.~(\ref{mixed}). Thus, these series just exhibit this intermediate 
behavior of the time correlations. Figures 6 and 7 display  
double-logarithmic plots of the normalized probability density of the  
$r_q$ values calculated for the fluctuations at $400$ K respectively of the  
resistance and of the defect-fraction. Different thresholds $q$ ranging  
from $2$ to $3.5$ are considered in both cases. The slope of the solid 
line is 0.26 and 0.23, respectively for the case of Figs. 6 and 7.   
Figures 6 and 7 show that the distribution of the return intervals 
of extreme values of these series is well described by a stretched  
exponential and that the value of the exponent $\gamma$ is independent  
of the threshold $q$ in a large range of $q$-values. This occurs even in 
absence of long-term correlations and in presence of a finite correlation 
time.  
 
\section{CONCLUSIONS}   
The distribution of return intervals of extreme values has been  
studied in several time series with different correlation properties: 
long-term and finite-term correlations. Precisely,  
it has been analyzed the return interval distribution of the fluctuations 
of resistance and defect-fraction displayed by a resistor with granular  
structure in nonequilibrium stationary states at different temperatures T. 
The resistance fluctuations were calculated by using the MSN model which 
is based on a resistor network approach \cite{pen_msn}. It has been found 
that when the auto-correlation function displays a non-exponential and a 
non-power-law decay, the distribution of the return intervals is well  
described by a stretched exponential with exponent $\gamma$ largely 
independent of the threshold $q$. This result shows that the stretched  
exponential distribution describes the distribution of the return  
intervals of extreme values not only when long-term correlations are present 
in the time series \cite{kantz_2005,bunde_physa2003,bunde_prl2005}, but even 
when finite-term correlations with non-exponential decay exist among the 
records, a situation typical of many systems which are approaching 
criticality. 
   
\acknowledgments   
Support from MIUR cofin-05 project "Strumentazione elettronica integrata 
per lo studio di variazioni conformazionali di proteine tramite misure  
elettriche" is acknowledged. The author thanks S. Ruffo (University of  
Florence, Italy), P. Olla (ISAC-CNR, Cagliari, Italy), G. Salvadori  
(University of Lecce, Italy) and E. G. Altmann (Max Planck Inst. for  
Phys. of Complex Systems, Dresden, Germany) for helpful discussions.

\end{document}